\documentclass[twocolumn,showpacs,pre,floatfix,longbibliography]{revtex4-1}

\usepackage{graphicx}
\usepackage{epstopdf}
\usepackage{epsfig}
\usepackage{amssymb,amsmath,stmaryrd,tabularx}
\usepackage{wrapfig}
\usepackage{bm}
\usepackage[center]{subfigure}

\newcommand{\dd}[2]{\frac{\mathrm{d}#1}{\mathrm{d}#2}}
\newcommand{\ddt}[2]{\frac{\mathrm{d^2}#1}{\mathrm{d}#2^2}}
\newcommand{\ddx}[2]{\ddt{}x}
\newcommand{\ddy}[2]{\ddt{}y}
\newcommand{\ddz}[2]{\ddt{}z}

\newcommand{\fv}[1]{\left\langle #1 \right\rangle} 
\renewcommand{\Re}{\operatorname{Re}}

\begin{document}
\title{The origin of the inverse energy cascade in two-dimensional quantum turbulence}
\author{Audun Skaugen and Luiza Angheluta} 
\affiliation{
  Department of Physics, University of Oslo, P.O. 1048 Blindern, 0316 Oslo, Norway
}
\pacs{47.27.-i, 03.75.Lm, 67.85.De}

\begin{abstract}
We establish a statistical relationship between the inverse energy cascade and 
the spatial correlations of clustered vortices in two-dimensional quantum turbulence. The Kolmogorov spectrum $k^{-5/3}$ on inertial scales $r$ corresponds to a pair correlation function between the vortices with different signs that decays as a power law with the pair distance given as $r^{-4/3}$. To test these scaling relations, we propose a novel forced and dissipative point vortex model that captures the turbulent dynamics of quantized vortices by the emergent clustering of same-sign vortices. The inverse energy cascade developing in a statistically neutral system originates from this vortex clustering that evolves with time. 
\end{abstract}
\maketitle

\section{Introduction}
The condensation of energy into large-scale coherent structures, and the prevailing flow of energy from small to large scales referred to as the inverse energy cascade~\cite{kraichnan1967inertial} are two important signatures of two-dimensional (2D) classical flows. The inverse energy cascade is associated with the statistical conservation laws of energy and enstrophy (mean-square vorticity). This is in stark contrast to three-dimensional turbulence where energy cascades towards small, dissipative length-scales by the proliferation of vorticity. Quantum turbulence (QT) in highly-oblate Bose Einstein condensates (BEC) provides a close experimental realization of 2D turbulence through the dynamics of quantized vortices, and a well defined theoretical framework to study turbulence from the statistical properties of a quantized vortex gas. In the recent years, there have been indirect experimental evidence~\cite{neely2013characteristics}, and tantalizing numerical results~\cite{Reeves_2012,Bradley_2012,Reeves_2013} of the existence of an inverse energy cascade that follows the Kolmorogov scaling analogously to the classical turbulence. The condensation of energy on large scales has also been investigated in decaying quantum turbulence~\cite{Simula2014, moon2015thermal, billam2015spectral}. A crucial link between these two phenomena is that there is a net transport of energy from small to large scales, and this happens in quantum turbulence due to the spatial clustering of quantized vortices of the same circulation. Here we show that this vortex interaction leading to clustering implies that the energy spectrum is directly related to spatial correlation functions, in particular to the vorticity correlation.  

Large-scale vortices were first predicted by Onsager~\cite{Onsager_1949} as the negative temperature equilibrium configuration of point vortices in a statistical description of 2D turbulence bounded in a finite domain. These vortex condensates are formed by clustering of vortices of the same sign, and occur through an SO(2) symmetry-breaking phase transition with negative critical temperature~\cite{yu2016theory}. Recent studies of decaying QT propose that such negative temperature states can be achieved dynamically by an evaporative heating process through the annihilation of vortex dipoles (effective heating of vortices by removing the coldest ones, i.e. the smallest dipoles)~\cite{Simula2014}. Clustering of vortices is also at the origin of an inverse energy cascade in \emph{driven QT}~\cite{Reeves_2013,Reeves_2012,skaugen2016vortex}, but in this case the mere existence of vortex clusters is not enough; their structure is crucially important and different from the equilibrium Onsager vortices emerging in decaying turbulence. Novikov~\cite{Novikov_1975} showed that, in principle, an inverse energy cascade with the Kolmogorov spectrum can be produced by a single cluster of same-sign vortices when there are long-range correlations in their spatial distribution, such that the pair distribution function decays as a power law $r^{-4/3}$ with the separation distance $r$. The dependence of the energy spectrum on the vortex configuration was further explored in this idealized case of same-sign point vortices~\cite{Bradley_2012,skaugen2016velocity}. However, any realistic model of driven QT needs to include both vortex signs, which complicates this idealized picture.

Our aim is to show that the inverse energy cascade in driven QT is the result of the interaction between diverse clusters of co-rotating and counter-rotating vortices. The largest of these clusters will be the nonequilibrium analogue of Onsager vortices, in the sense that they will dominate the large-scale rotating flow. However, there is a spectrum of vortex clusters of various sizes, and their interactions and internal structures lead to persistent, long-range fluctuations in the vorticity field, measured by the weighted pair correlation function. Moreover, we show that the Kolmogorov spectrum is related to scale-free two-point statistics of quantized vorticity, $\fv{\omega(0)\omega(\vec r)}\sim r^{-4/3}$, similar to Novikov's prediction for the idealized same-sign vortices. This would be analogous to the Kraichnan-Kolmogorov scaling argument for the classical coarse-grained vorticity on an inertial scale $r$, $\omega_r\sim v_r/r$, with the eddy velocity $v_r\sim r^{1/3}$~\cite{kraichnan1967inertial}. To investigate the origin of the inverse energy cascade in driven quantum turbulence without having to concern ourselves about the compressibility effects in quantum fluids~\cite{neely2013characteristics}, we propose a driven and dissipative point vortex model as described below. 

The structure of the paper is as follows. In Section II we describe the point vortex model as applied to quantum turbulence. In section III we give a statistical argument relating the energy spectrum to the 
vorticity correlation. This connection is further explored numerically within a new driven and dissipative point vortex model which is introduced in Section IV. We show that this model develops a turbulent steady-state in Section V. In 
section VI, we discuss the numerical results on the vorticity correlation function, and in Section VII we show that this is associated with a negative spectral energy flux and a Kolmogorov $k^{-5/3}$ energy spectrum. Final conclusions and discussion are presented in Section VIII.


\section{Point vortex model}
Point vortices in BECs are realized as codimension two topological defects in the complex order parameter representing the many-particle wave function. These vortices have a characteristic 
core structure determined by the balance of kinetic energy and distortion energy, leading to a core size on the order of the healing length $\xi$~\cite{pismen1999,Bradley_2012}. 
Vortices with overlapping cores have complicated dynamics which couple the vortex gas to phonon excitations in the BEC, and lead to the annihilation of vortex-antivortex
pairs. On the other hand, well-separated vortices will move according to the point vortex model. 
A collection of vortices with circulation signs $\{q_i\}$ and positions $\{\vec r_i\}$ in a bounded container of size $R$ set up a velocity field characterized by the stream function $\psi(\vec r,t)$, given by a superposition of contributions from each vortex~\cite{newton2013n}
\begin{equation}
  \psi(\vec r,t) = -\sum_{j=1}^N q_j \ln|\vec r-\vec r_j| + \sum_{j=1}^N q_j\ln(|\vec r - \vec r_j^{\, v}|r_j).
\end{equation}  
The second sum gives the contributions from image vortices located at $\vec r_j^{\, v} = \frac{R^2}{r_j^2}\vec r_j$, which are necessary in order to
ensure the no-flux boundary condition at $r = R$, as well as image vortices located at the origin giving the extra factor of $r_j= |\vec r_j|$ in the logarithm. 
Note that the winding number $q_j = \pm 1$ is constant for quantized vortices in 2D QT, unlike the continuously varying circulation of classical vortices subjected to merging rules in vortex models for 2D CT~\cite{benzi1992simple}. Well separated vortices follow this velocity field passively,
\begin{equation}
  \dot{\vec r}_i = \nabla^\bot \psi^{(i)}(\vec r_i), 
\end{equation}
where $\nabla^\bot = (-\partial_{y},\partial_{x})$ and the $(i)$ superscript indicates that we omit the singular self-interaction from the $i=j$ term of the streamfunction (although we keep the $i=j$ term in the image sum, which gives a non-singular interaction between a vortex and its image). 
This is equivalent to a conservative dynamics described by the Hamiltonian 
\begin{align}
  \mathcal{H} &= \frac 1 2 \sum_i q_i\psi^{(i)}(\vec r_i) \label{eq:H} \\
  &= -\frac 1 2 \sum_{i\ne j} q_iq_j\ln|\vec r_i-\vec r_j| + \frac 1 2 \sum_{i,j}q_iq_j \ln (|\vec r_i - \vec r_j^{\, v}|r_j), \notag
\end{align}
reflecting the conservation of kinetic energy in the velocity field. 

Although we can always make sure that the initial vortex positions are well separated from each other, the dynamical evolution might cause pairs of vortices to get close enough that the coupling to the phonon
field becomes important. It is therefore necessary to include phenomenological rules such as dipole annihilation in order to properly represent BEC dynamics in a point vortex model.

The Hamiltonian point vortex model, describing conservative dynamics, cannot capture dynamical aspects of turbulence such as the energy cascade or the build-up of large scale coherent structures. It is however useful for investigating equilibrium statistical properties for inertial turbulent fluctuations~\cite{pointin1976statistical}. The phase space of the point vortex Hamiltonian coincides with the configuration space, hence is finite for a bounded domain. This implies that the microcanonical entropy has a maximum at finite energy, giving rise to negative temperature $T$ states at higher energies. Onsager~\cite{Onsager_1949} predicted that these negative $T$ states correspond to spatial clustering of same-sign vortices, resulting in large-scale vortices. Recently, it has been proposed that the Onsager vortices in the strongly coupled regime undergo a condensation similar to the Bose-Einstein condensation, but this requires much higher energies that those present in decaying turbulence or in the inverse energy cascade~\cite{valani2016einstein}.

For dynamical exploration of these vortex clustered configurations from initial conditions with positive $T$, one needs to include phenomenological dissipation mechanisms, such as dipole annihilation rules~\cite{Simula2014} and/or thermal friction~\cite{kim2016role}. In Ref.~\cite{campbell1991statistics}, it was proposed that the annihilation of the smallest vortex dipoles acts as an effective viscous dissipation in QT. However, as pointed out in Ref.~\cite{Simula2014} this~\lq effective viscosity\rq~is not constant, but instead depends on the spatial configuration of vortices and vanishes for clusters of same-sign vortices. By removing the smallest vortex dipoles, the total energy will slowly decrease with decreasing number of vortices. Since the energy of the smallest dipole is smaller than the mean energy per vortex, the net energy per vortex keeps increasing. Hence it acts like an evaporative heating mechanism and leads to the emergence of same-sign vortex clusters in decaying turbulence~\cite{Simula2014}.

\section{Energy spectrum and vorticity correlation}
The kinetic energy spectrum of $N$ point vortices is determined by their spatial configuration and charges as derived by Novikov~\cite{Novikov_1975} and in more recent studies of BEC vortices with a characteristic vortex core structure~\cite{Bradley_2012}. For a given vortex configuration, the energy spectrum of $N$ vortices in an unbounded plane is given as~\cite{Novikov_1975}    
\begin{eqnarray}
  E_N(k) = \frac{\pi}{k}\left( N + \sum_{i \neq j}q_iq_j J_0(kr_{ij}) \right) \label{eq:espec},
\end{eqnarray}
where $J_0(x)$ is the zeroth order Bessel function and $r_{ij} = |\vec r_{ij}| = |\vec r_i - \vec r_j|$. To study this statistically, Novikov considered a cluster of same-sign vortices and introduced the pair correlation 
$g(\vec r) = \frac{1}{\rho\fv N}\fv{\sum_{i\neq j}\delta(\vec r - \vec r_{ij})}$ with the vortex number density
$\rho = \fv{\sum_i\delta(\vec r - \vec r_i)}$. For vortices inside the given cluster the sign factors $q_iq_j$ will equal $1$, so averaging the kinetic energy spectrum we find
\begin{eqnarray}
  \fv{E_N(k)} &=& \frac{\pi}{k}\left( \fv{N} + \int J_0(kr)\fv{\sum_{i\neq j}\delta(\vec r - \vec r_{ij})}d^2 \vec r \right) \notag\\
  &=& \frac{\fv N \pi}{k}\left( 1 + \rho\int J_0(kr)g(\vec r)d^2\vec r \right).
  \label{eq:novcorr}
\end{eqnarray}
A scaling $g(\vec r) \sim 1 + Cr^{-\alpha}$ will then give rise to two new terms $\sim\delta(k)$ and $\sim k^{\alpha-3}$ in the energy spectrum, in addition to the $\sim k^{-1}$ term from the single-vortex solution. This is Novikov's statistical relationship between the scalings in energy and pair correlation of the same-sign vortex gas. 

However, even if the vortices organize into clusters with a characteristic pair correlation $g(r)$, it is not clear that the clusters would be decoupled sufficiently from each other for energy spectrum to be a simple superposition of the spectra of each cluster. We therefore argue that simply performing the average inside a given vortex cluster does not give the complete statistical description of turbulence with two vortex signs. Instead, one should perform this derivation accounting for all the different vortex signs.

Averaging equation~(\ref{eq:espec}) and keeping all the sign factors, we find
\begin{eqnarray}\label{eq:Ek}
  \fv{E_N(k)} &=& \frac{\pi}{k}\left( \fv{N} + \fv{\sum_{i\neq j}q_iq_j J_0(kr_{ij})} \right) \notag \\
  &=& \frac{\pi}{k}\left( \fv{N} + \int d^2\vec r J_0(kr)\fv{\sum_{i\neq j}q_iq_j \delta(\vec r - \vec r_{ij})} \right) \notag \\
  &=& \frac{\fv{N}\pi}{k}\left( 1 + \int d^2\vec r J_0(kr) \rho g_w(\vec r) \right),
\end{eqnarray}
where the weighted pair correlation $\rho g_w(\vec r)$ is defined as 
\begin{eqnarray}\label{eq:gw}
  \rho g_w(\vec r) &=& \frac{1}{\fv{N}}\fv{\sum_{i \ne j} q_iq_j \delta(\vec r-\vec r_i+\vec r_j)}. 
\end{eqnarray}
This function can be interpreted as the probability of finding two same-sign vortices at a separation $\vec r$, minus the probability of finding two opposite-sign vortices at the same separation. 
For vortices of same sign, we recover the simple pair correlation $\rho g(\vec r)$, which decreases monotonically such that $\lim_{r\rightarrow\infty}g_w(r) =1$, reflecting that the vortex positions are uncorrelated at large distances. 
But for the neutral system the two probabilities will cancel out, giving $\lim_{r\to\infty}g_w(r) = 0$, reflecting the fact that there should be no excess of one sign over the other at large distances.

At intermediate distances the $g_w(r)$ function indicates the predominance of vortex clusters, giving positive values, and vortex dipoles, giving negative values. A characteristic scale might indicate the typical size of clusters, while
a scale-free behavior might indicate either the coexistince of clusters of different sizes, or a scale-free spatial structure of each cluster. Similarly to Novikov's argument, the Kolmogorov $k^{-5/3}$ energy scaling corresponds to
a $Cr^{-4/3}$ scaling in the weighted pair correlation, by the relation
\begin{eqnarray}\label{eq:EkKol}
  \frac{\fv{E(k)}}{\fv{N}} \sim \frac{\pi}{k} + 2\pi^2\frac{\rho C}{k^{5/3}} \int dx J_0(x) x^{-1/3}.
\end{eqnarray}
The vanishing limit of the weighted pair correlation means that there is no singularity $\sim\delta(k)$ in the energy spectrum. The dimensionless integral equals $\frac{\sqrt \pi}{\Gamma(5/6)}$ in an unbounded system, but may introduce
corrections in a finite-size system.

The weighted pair correlation function is related to correlations in the vorticity field $\omega(\vec r) = \sum_{i}q_i\delta(\vec r - \vec r_i)$. Assuming a statistically homogeneous system, two-point correlations in vorticity can
be found as
\begin{eqnarray}
  \fv{\omega(\vec r)\omega(\vec r')} &=& \fv{\sum_{i,j} q_iq_j\delta(\vec r - \vec r_i)\delta(\vec r'-\vec r_j)} \notag\\
  &=& \rho\delta(\vec r - \vec r') + \fv{\sum_{i\neq j}q_iq_j \delta(\vec r - \vec r_i)\delta(\vec r'-\vec r_j)} \notag\\
  &=& \rho\delta(\vec r - \vec r') + \rho^2 g_w(\vec r - \vec r'). 
  \label{eq:vortcorr}
\end{eqnarray}
Hence the $r^{-4/3}$ scaling in the weighted pair correlation relates to a similar scaling in the vorticity correlation. Such a scaling in the vorticity is analogous to the Kraichnnan-Kolmogorov scaling in classical turbulence, which follows from dimensional analysis based on statistical self-similarity of turbulence. 

\section{Driven and dissipative point vortex model}
In order to numerically investigate the relationship between energy spectrum and vorticity correlations, we propose an extension to the point vortex model which can capture driven quantum turbulence, by adding driving and dissipation 
mechanisms to the equations of motion.

A natural source of dissipation in QT is the non-conservative vortex motion due to thermal friction. This adds a longitudinal component in the vortex motion, with repulsion between same-sign vortices and attraction between opposite-sign vortices. Hence, the point vortices follow a non-Hamiltonian equation of motion given as 
\begin{equation}\label{eq:DPV}
  \dot{\vec r}_i = \nabla^\bot \psi^{(i)}(\vec r_i)-\gamma q_i\nabla\psi^{(i)}(\vec r_i), 
\end{equation}
where $\gamma$ is the dimensionless thermal friction coefficient, and the sign factor $q_i$ is necessary in order to give different motion for same- and opposite-sign vortices. As discussed in Ref.~\cite{kim2016role}, $\gamma$ quantifies the main source of dissipation in the system and is directly related to the damping coefficient commonly used in Gross-Pitaevskii dynamics to account for thermal dissipation~\cite{Bradley_2012,Reeves_2012}. 

This dissipative evolution will cause vortex-antivortex pairs to collapse into ever tighter dipoles. To avoid a singularity we introduce a phenomenological annihilation rule for such pairs when the distance is closer than 
a constant $l_a$. This constant can for example represent a length scale on the order of the healing length $\xi$ in a BEC. Vortices in a bounded disk will also be attracted to the boundary, so we need a similar rule causing vortices close to the boundary to annihilate with their image. Although same-sign vortex pairs also behave differently in BECs when they get close enough together, we do not 
add any phenomenological rules for this case, as we do not expect any such pair to come within a distance much smaller than $l_a$ in our dynamics.

\begin{figure}[t]
  \centering
	\includegraphics[width=0.5\textwidth]{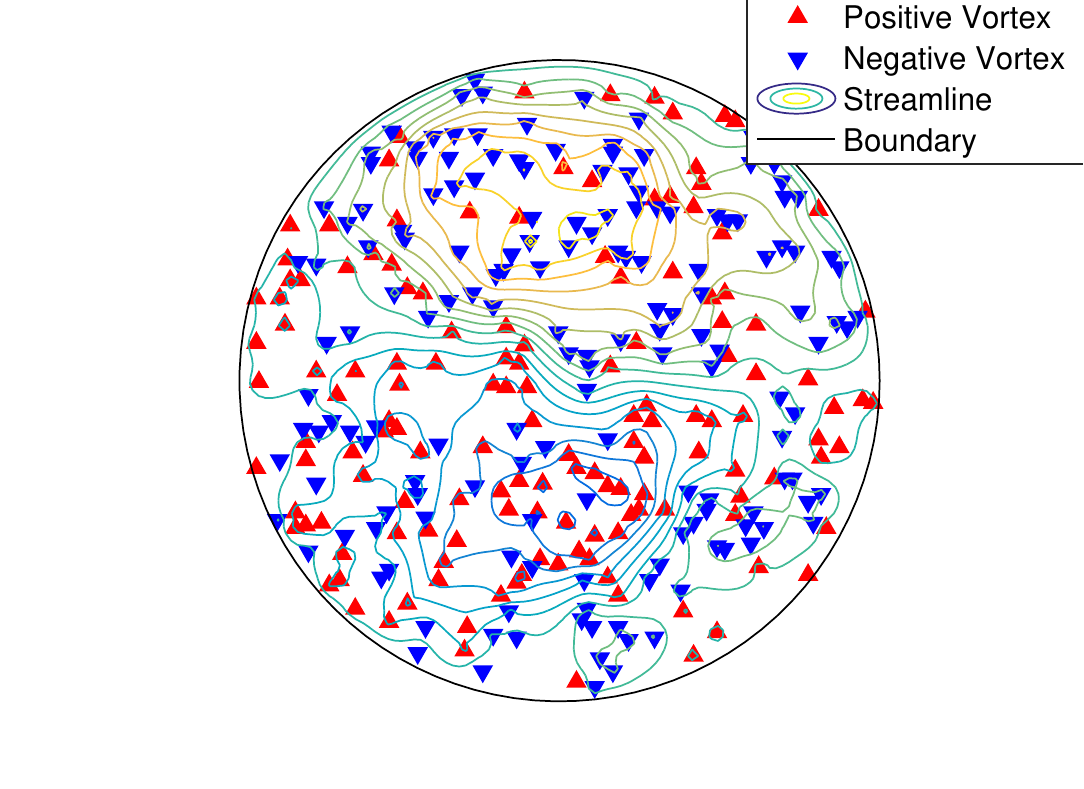}
  \caption{Snapshot of a vortex configuration.}  \label{fig:Snap}
\end{figure}

To generate forced turbulence, we include a small-scale stirring mechanism by spawning vortex-antivortex dipoles with a fixed separation $l_s>l_a$. This is different from the stirring mechanism with an effective negative-viscosity proposed by the Siggia and Aref~\cite{Siggia} and more relevant for QT. Our stirring mechanism is consistent with the setup for 2D QT in BECs~\cite{Reeves_2012,Reeves_2013,skaugen2016vortex}, where a moving obstacle causes dipoles to be spawned into the system at a distance $l_s$ bigger than the coherence length~$\xi$, and with annihilation occurring on the scale $\xi$ due to phonon emission from the Gross-Pitaevskii dynamics. When the dipole spawning and evaporation happen at equal rates, the total energy will increase roughly by $\ln(l_s/l_a)$ with each event. The increase in mean energy per vortex causes the spawned dipoles to decouple into free vortices, which then form clusters of same-sign vortices. Moreover, this driving occurs on a length-scale $\sim l_s$, so by choosing this distance to be small compared to the system size, we can achieve small-scale stirring. In the Supplemental Material~\cite{supplemental}, we have included two movies with the vortex dynamics during the transition to the steady state versus that in the statistically stationary turbulent regime.

We solve the system of ordinary differential equations given by Eq.~(\ref{eq:DPV}) using the symplectic, fully implicit $4$th order Gauss-Legendre method for the conservative part~\cite{mclachlan1992accuracy}. After each timestep of the symplectic method, we evolve the dissipative part using a simple forward Euler scheme, which is sufficient because $\gamma \ll 1$, so the dissipative evolution is slow compared to the conservative evolution. We use an adaptive timestep based on the minimum distance $d$ between vortices and vortex-image pairs. Since the maximum velocity is $v \sim 1/d$, the typical time until a collision $\Delta t \sim d / v \sim d^2$ is a reasonable choice for the adaptive timestep. With each timestep increment, we remove all vortex dipoles with dipole moments $<l_a$, and any vortex within a distance $<l_a$ from its image vortex. Finally, we pick a random number $n$ from a Poisson distribution with the rate parameter $c \Delta t$ (with $c$ being the dipole injection rate), and spawn $n$ dipoles with dipole moment $l_s$ at random positions and dipole orientations, such that neither of the spawned vortices is within $l_s$ of another vortex. 

During these simulations, we fixed $R = 10$, $l_s = 0.4$ and $l_a = 0.2$. The thermal friction coefficient $\gamma$
and the spawning rate $c$ were varied to generate different regimes of turbulence. The values are given in Table
\ref{tab:params}, along with measured values for the mean vortex number $\fv{N}$ with the resulting Reynolds number (see Section \ref{sec:steady}),
and the time-averaged total energy dissipation $\fv{\epsilon_d^{\text{tot}}}$ (see Section \ref{sec:spec}).

\begin{figure}[t]
  \centering
  \includegraphics[width=0.5\textwidth]{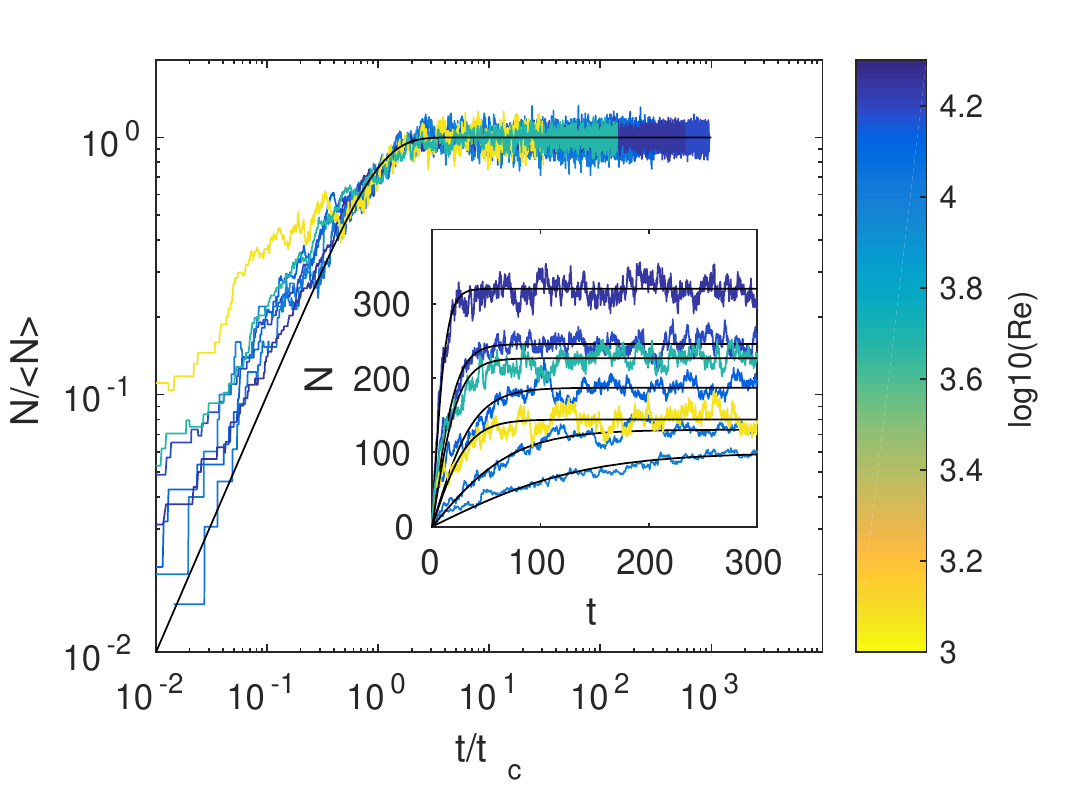}
  \caption{Inset: Temporal evolution of the total vortex number for different $\Re$ numbers. We extract the mean number $\fv{N}$ and the cross-over time $t_c$ by fitting each curve with the $\tanh$ function. Main figure: The collapsed data in rescaled units plotted against $\tanh(x)$.}  \label{fig:VortexNum}
\end{figure}

\section{Steady state turbulent regime}
\label{sec:steady}
As a consequence of spawning and evaporation of vortex dipoles, the total number of vortices fluctuates in time. Starting from zero vortices, the number of vortices increases by spawning events dominating over dipole evaporation events. After a transient time, the number fluctuation reach a statistically steady state regime as shown in Figure~\ref{fig:VortexNum} for different $\Re$ numbers. By a rescaling of time with the cross-over time $t_c$ and of number fluctuations with their mean value $\fv{N}$, we find that all data collapse onto an universal curve that is nicely fitted by the $\tanh(x)$ function, i.e. the solution of the mean field kinetic equation $\dot N = -N^2+1$, written in rescaled units of $t/t_c$ and $N/\fv{N}$. The first term is the leading order contribution due to dipole annihilation, and the last term is the constant spawning rate. Small deviations from this mean-field trend are the effect of higher order terms due to collective interactions in dipole evaporation~\cite{groszek2016onsager}, but also because we neglected the small effect of a linear decay term corresponding to the evaporation of single vortices near the boundary of the disk. 

\begin{table}
  \centering
  \caption{Parameters used in the simulations, along with some measured quantities.}
  \begin{tabular}{r|r|r|r @{.} l|r @{.} l}
	\hline\hline
	\multicolumn{2}{c|}{Parameters} & \multicolumn{5}{c}{Measured quantities}\\
	$10^3\gamma$ & $c$ & $\fv{N}$ & \multicolumn{2}{c|}{$10^{-3}\Re$} & \multicolumn{2}{c}{$\fv{\epsilon_d^{\text{tot}}}$} \\ \hline
	10 & 10 & 144 & 1&20 &  6&59 \\
	 3 & 10 & 227 & 5&02 &  5&77 \\
	 1 &  1 & 100 & 10&0 &  0&541 \\
	 1 &  2 & 131 & 11&4 &  1&13 \\
	 1 &  5 & 187 & 13&7 &  2&79 \\
	 1 & 10 & 246 & 15&7 &  5&33 \\
	 1 & 20 & 320 & 17&9 & 10&5 \\ \hline \hline
  \end{tabular}
  \label{tab:params}
\end{table}

After the vortex number has stabilized, the energy of the system (as measured by Eq.~\ref{eq:H}) keeps increasing as dipoles decouple into free vortices which then form clusters. 
Higher energy leads to a higher energy dissipation rate from the dissipative evolution, eventually balancing the small-scale forcing, so the energy eventually also stabilizes.  By carefully balancing the spawning and dissipation rates we can make sure that the steady-state energy is such that the system is dominated by clusters.

For the statistically stationary regime (a snapshot of the vortex configuration is illustrated in Figure~\ref{fig:Snap}, and a video corresponding to this regime is included in the Supplemental Material~\cite{supplemental}), we can define a Reynolds (Re) number by the balance between inertial forces (where the typical distance is the disk's radius $R$ and the typical field velocity is $U\sim 1/\fv{l}$ with $\fv{l}\sim R/\sqrt{\fv{N}}$) and dissipative forces (thermal friction characterized by $\gamma$), hence $\Re \sim \sqrt{\fv{N}}/\gamma$.  

\begin{figure}[t]
  \begin{center}
    \includegraphics[width=0.3\textwidth]{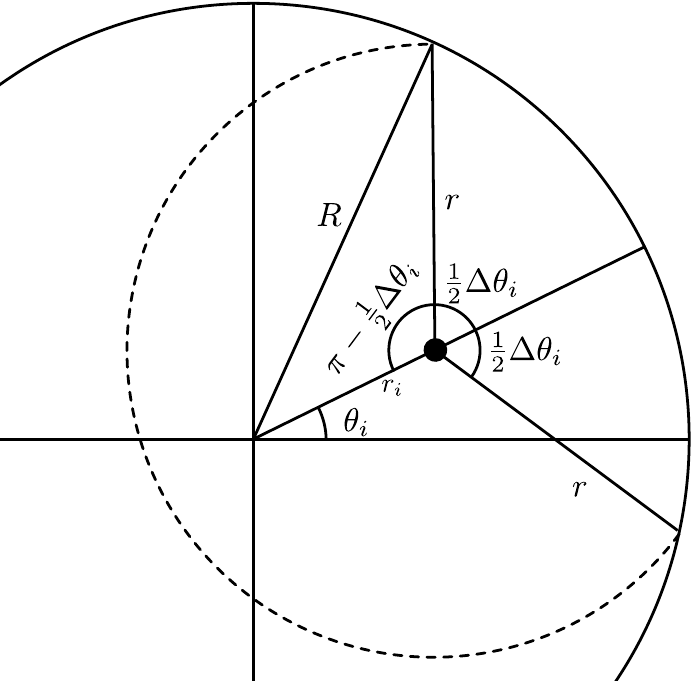}
    \caption{Directions not crossing the boundary for a given vortex at $\vec r_i$ (filled circle). The solid circle gives the boundary, while the dashed circle gives the shell of radius $r$.}
    \label{fig:shell}
  \end{center}
\end{figure}
\section{Weighted pair correlation function in a disk}
In a statistically homogeneous and isotropic system, the weighted pair correlation function defined in Eq.~(\ref{eq:gw}) is also isotropic, and therefore is equivalent to 
\begin{align}
\rho g_w(r) &= \frac{1}{2\pi r}\int_0^{2\pi} \rho g_w(\vec r)r d\theta\nonumber\\
&= \frac{1}{2\pi r \fv{N}}\sum_{i\neq j}\fv{q_iq_j\delta(r - r_{ij})},
\end{align}
where $r_{ij} = |\vec r_i - \vec r_j|$. For a numerical estimate of this expectation value, we can discretize the delta function into bins of width $d$, by
$\delta_d(r) = [H(r + \frac{d}{2}) - H(r-\frac{d}{2})]/d$, where $H(x)$ is the step function. This gives the contribution $g_w^s(r)$ to the correlation function from a given realization as 
\begin{align}
  \fv{N}\rho g_w^s(r) = \frac{1}{2\pi r d}\sum_{i\neq j}q_iq_jd\delta_d(r-r_{ij}).
\end{align}
%

\begin{figure}[t]
  \centering
	\includegraphics[width=0.5\textwidth]{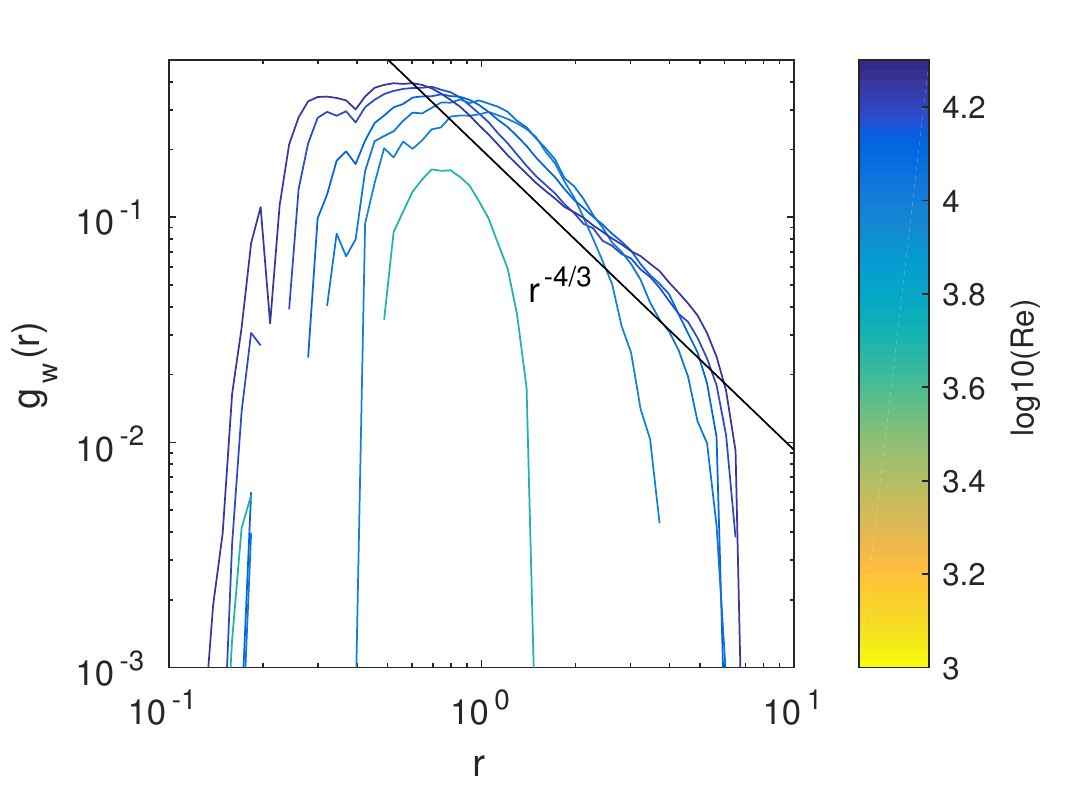}
  \caption{Weighted pair correlation function for different Re numbers, ignoring isolated vortices. The case with the lowest Re number $\Re \sim 1000$ is excluded because this system is dominated by dipoles, so that $g_w(r) < 0$ in most of its range.}
  \label{fig:radfs}
\end{figure}

Thus we can estimate the weighted correlation function by iterating over each vortex, counting the number of vortices which fall within a shell of radius 
$r$ and width $d$, weighting them by whether they have equal or opposite sign, and dividing by the area of the shell. By time-averaging over many vortex configurations in the statistically stationary regime, we can divide by the mean vortex number and density, giving an estimate of the correlation function.

A complication to this method arises from the fact that our system is finite with a circular boundary of radius $R$. This breaks homogeneity and isotropy, 
especially when considering particles close to the boundary. We will however still assume that the system is as homogeneous and isotropic as possible, 
by which we mean the following: Considering a vortex close to the boundary, the system is assumed to look the same in all directions as it would look from the
center, as long as we do not see the boundary. Taking the contribution from each vortex separately, 
\begin{align}
\fv{N}\rho g_w^s(\vec r) &= \sum_i q_i g_w^i(\vec r),\nonumber\\
 g_w^i(\vec r) &= \sum_{j\neq i}q_j \frac 1 r \delta(r-r_{ij})\delta(\theta-\theta_{ij}), 
\end{align}
we assume that $g_w^i(\vec r)$ is equal for all directions $\theta$ such that $\vec r + \vec r_i$ does not cross the boundary. 
\begin{figure}[t]
  \centering
	 \includegraphics[width=0.5\textwidth]{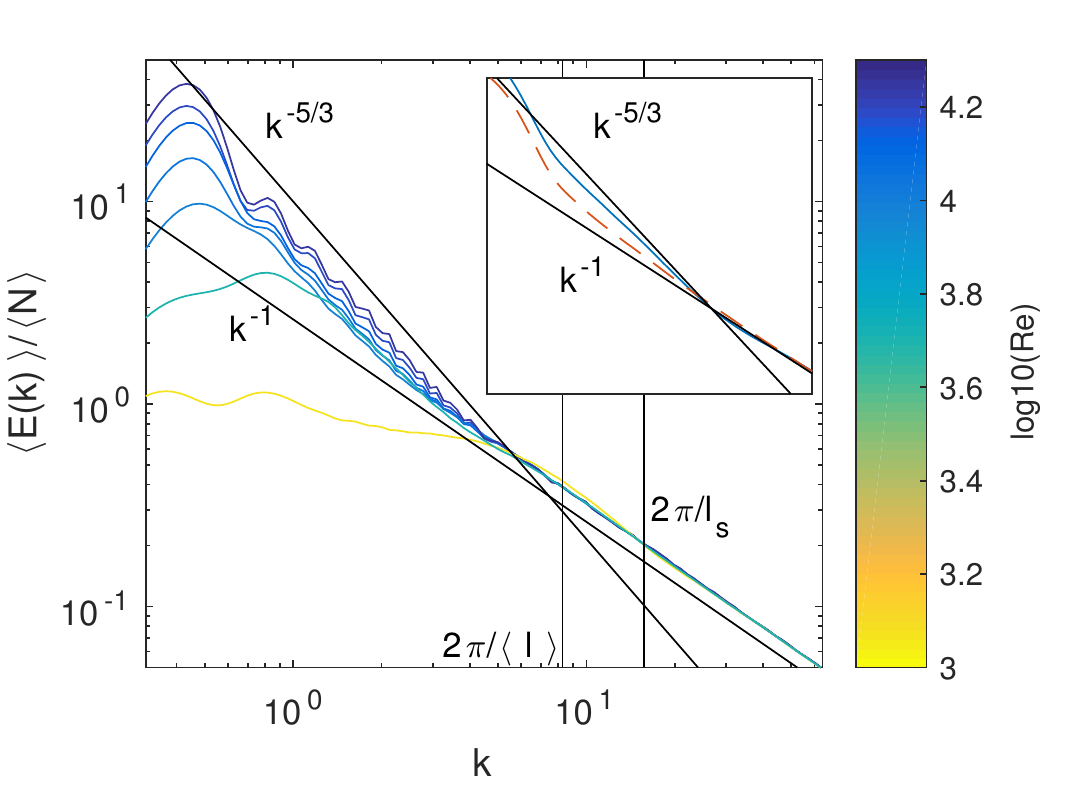}
  \caption{Energy spectrum per vortex for different $\Re$ numbers, ignoring single vortices. The vertical lines correspond to a typical inter-vortex spacing $\fv{l}$ and the spawning lengthscale $l_s$. In the inset we compare two different energy spectra for the highest $\Re$ number, where we ignore the contribution from the image vortices: Including all vortices (dashed line), and excluding isolated vortices (solid line). Clearly, excluding isolated vortices is necessary to show Kolmogorov scaling.}
  \label{fig:specs}
\end{figure}

These directions go from $\theta_i^+ = \theta_i + \frac 1 2 \Delta \theta_i$ to $2\pi + \theta_i^- = 2\pi + \theta_i - \frac 1 2 \Delta \theta_i$ 
(as illustrated in Figure \ref{fig:shell}), and by the law of cosines we have 
\begin{align}
  R^2 &= r^2 + r_i^2 + 2rr_i\cos \left(\frac 1 2 \Delta\theta_i\right), \nonumber\\
  \Delta\theta_i &= 2\arccos \frac{R^2 - r^2 - r_i^2}{2r r_i}.
\end{align}
By our assumption of near isotropy we can integrate over the allowed directions, 
\begin{align}
  g_w^i(r) &= \frac{1}{(2\pi-\Delta\theta_i)r}\int_{\theta_i^+}^{2\pi + \theta_i^-}g_w^i(\vec r)r d\theta \nonumber\\
  &= \frac{1}{(2\pi-\Delta\theta_i)r} \sum_{j\neq i}q_j\delta(r - r_{ij}), 
\end{align}
where we used that $\theta_{ij}$ certainly lies within the integration range if $r = r_{ij}$, otherwise $\vec r_j$ would be outside the boundary. 

Thus we can account for the boundary effects by estimating the correlation function in the same way as for an unbounded system, only reducing the area of the shell of radius $r$ around vortex $i$ from $2\pi rd$ to the area given
by the directions not crossing the boundary, \[A_i = \left(2\pi - 2\arccos \frac{R^2-r^2-r_i^2}{2rr_i}\right)rd. \]

We measured $g_w(r)$ taking into account this boundary effect, and time-averaging over $10^3-10^4$ statistically-stationary realizations outputted at regularly spaced time intervals (about the typical timescale it takes an injected dipole to cross the disk). The resulting distribution is shown for different $\Re$ numbers in Figure~\ref{fig:radfs}, where we removed the spurious contributions of single (unclustered) vortices, using the same clustering analysis as in Ref.~\cite{skaugen2016vortex}. For the smallest $\Re$ number, $g_w(r) < 0$ for most of its range, indicating that the system is dominated by dipoles. At larger $\Re$ numbers it develops a peak around a characteristic cluster size $r_c$ and falls off rapidly after this, which indicates that the system is dominated by small vortex clusters with similar sizes. However, at sufficiently large $\Re$ number, we observe instead a scaling regime developing and approaching the power-law $C r^{-4/3}$, indicating a diverse range of different cluster sizes.

\section{Energy spectrum in a disk}
\label{sec:spec}

For $N$ vortices, the energy spectrum from Eq.~(\ref{eq:espec}) can be computed in $O(N^2)$ steps straightforwardly. However, with the imposed circular boundary at radius $R$, there are additional contributions due to image vortices. With these contributions worked out in Ref.~\cite{yoshida2005numerical}, the energy spectrum is given as
\begin{widetext}
\begin{align}
  E(k)= \frac{\pi}{k}\Bigg\{N + 2\sum_{i<j}q_iq_j J_0(kr_{ij})
  &+ \sum_{l=0}^\infty \epsilon_l J_l(kR)\sum_{i,j}q_iq_j \left( \frac{r_i}{R} \right)^l\Bigg[ J_l(kR)\left(\frac{r_j}{R}\right)^l 
	- 2J_l(kr_j) \Bigg]\cos(l\theta_{ij}) \Bigg\},
\end{align}
\end{widetext}
where $\epsilon_0 = 1$, $\epsilon_l = 2$ for $l \ge 1$, and $\theta_{ij} = \theta_i - \theta_j$ is the angle between vortices $i$ and $j$.

Numerically, these extra sums are expensive to compute because they involve $N^2 L$ terms, where $L$ is the number of terms we use in the $l$ summation. 
However, the extra terms can be transformed into products of single sums over the number of vortices, decreasing the cost to $NL$, and making it simpler to implement. 

The key insight is that both the finite-size sums can almost be factored into independent sums over $i$ and $j$, if not for the cosine terms coupling them. This allows us to split the sum into two new sums which can then be decoupled, 
\begin{widetext}
  \begin{align}
	E(k) = \frac{\pi}{k}\Bigg\{N + 2\sum_{i<j}q_iq_j J_0(kr_{ij})
	+ \sum_{l=0}^\infty \epsilon_l J_l(kR)\Bigg[\sum_i a_{il}\cos(l\theta_i)\sum_jb_{jl}\cos(l\theta_j)
	+ \sum_ia_{il}\sin(l\theta_i)\sum_jb_{jl}\sin(l\theta_j)\Bigg]\Bigg\}, \label{eq:especsums}
  \end{align}
\end{widetext}
where 
\begin{align}
  a_{il} = q_i \left( \frac{r_i}{R} \right)^l, \ b_{il} = q_j\left[ J_l(kR)\left( \frac{r_j}{R} \right)^l - 2J_l(kr_j) \right].
\end{align}
Here all the finite-size sums over $N$ vortices are independent, giving much faster numerical evaluation.

In Figure~\ref{fig:specs}, we show the energy spectrum per vortex $\fv{E(k)}/\fv{N}$ for the different $\Re$ numbers, time-averaged in the same way as with the correlation function. We notice that the energy spectrum transitions from the $1/k$ scaling attributed to the self-energy of a single vortex to the Kolmogorov $-5/3$ scaling law on wavenumbers smaller that $2\pi/\fv{l}$ at high $\Re$ numbers. As seen from Eq.~(\ref{eq:EkKol}), both contributions will be present in the energy spectrum, and even though the $-5/3$ contribution should in principle dominate the scaling at low $k$, the finite size effects may prevent this. 

In order to clearly see this inertial scaling regime both for $g_w(r)$ and $\fv{E_N(k)}$, we removed from the statistical analysis the contribution of single (unclustered) vortices, which can hinder the collective effects for small system sizes. The inset of Figure ~\ref{fig:specs} shows the spectral analysis for the highest $\Re$ number, when we include or exclude the effects of single vortices. The best inertial scaling seems to follow when we take out these effects and only look at the spectrum induced by clustered vortices. Including the contribution of the vortex images into the statistical analysis did not significantly change this picture, but merely introduced some minor oscillations around the trend line, which are visible in main part of Figure~\ref{fig:specs}. The presence of an inverse energy cascade indicates that the energy dissipation introduced in Eq.~(\ref{eq:DPV}) mostly acts on large scales, so that energy needs to transfer from the small injection scale to the larger dissipation scale. We now turn to measuring whether this is true.

\subsection*{Spectral energy flux and dissipation rate}

\begin{figure*}[t]
  \begin{center}
	\includegraphics[width=0.49\textwidth]{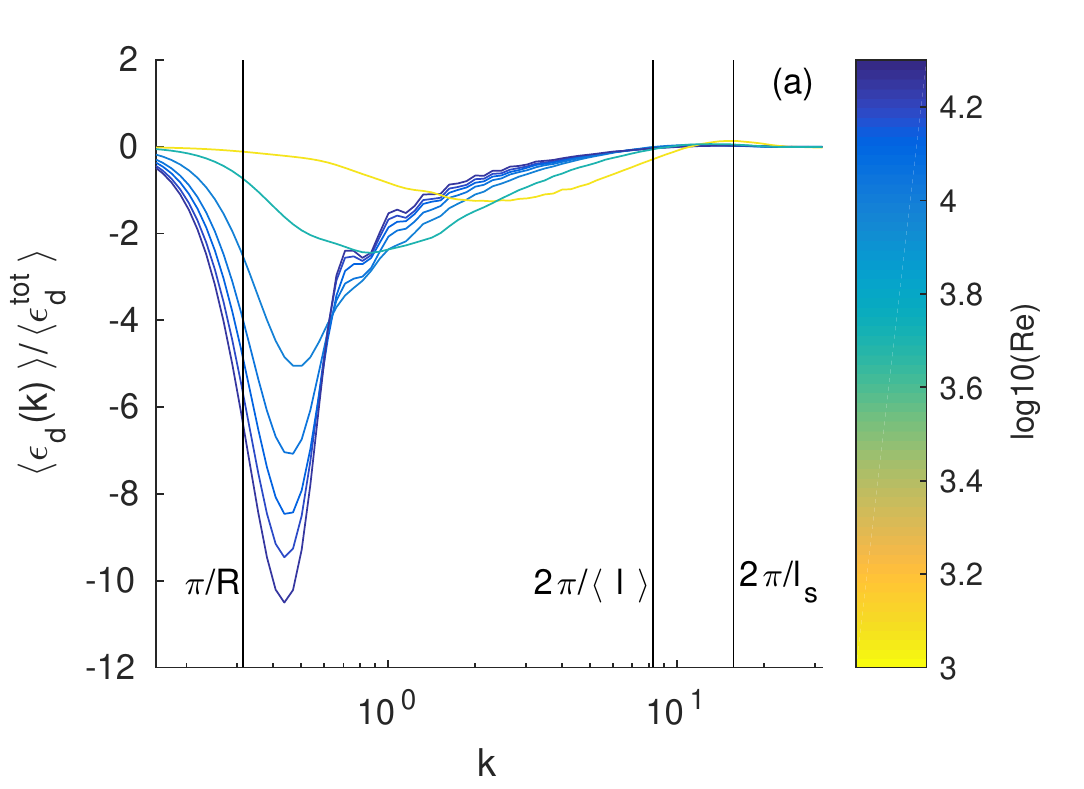}
	\includegraphics[width=0.49\textwidth]{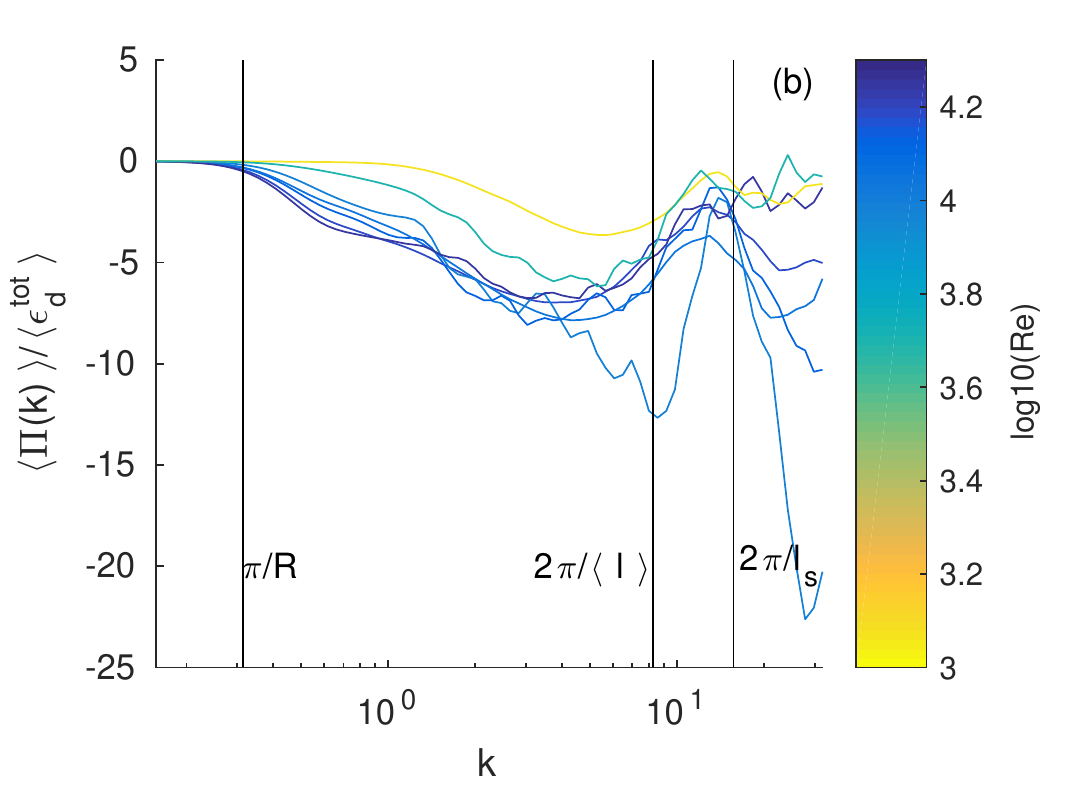}
  \end{center}
  \caption{(a) Normalized spectral dissipation rate and (b) spectral flux for different Reynolds numbers. At high $\Re$ the dissipation rate concentrates at large scales. The spectral flux is negative in a the inertial range, especially
  for high $\Re$.}
  \label{fig:fluxes}
\end{figure*}
The energy fluxes can be found by time-differentiating the energy spectrum from Eq.~(\ref{eq:espec}), giving
\begin{align}
  \dd{E(k)}{t} = -2\pi\sum_{i<j}q_iq_j J_1(kr_{ij})\frac{\vec r_{ij}}{r_{ij}}\cdot(\vec v_i - \vec v_j). \label{eq:flux}
\end{align}
Splitting the velocities into a dissipative part and a conservative part, $\vec v_i = \vec v_i^c + \vec v_i^d$, we have that 
\begin{align}
  \dd{E(k)}{t} = &-2\pi\sum_{i<j}q_iq_j J_1(kr_{ij})\frac{\vec r_{ij}\cdot \vec v_{ij}^d}{r_{ij}} \nonumber\\
	&- 2\pi\sum_{i<j}q_iq_j J_1(kr_{ij})\frac{\vec r_{ij}\cdot \vec v_{ij}^c}{r_{ij}}\nonumber\\
	= &\epsilon_d(k) + \epsilon_c(k). \label{eq:fluxes}
\end{align}
The second term is due to conservative evolution, and can therefore only serve to redistribute energy in the spectrum, $\int_0^\infty \epsilon_c(k) dk = 0$. 
Integrating this up to a given wavenumber $k$ gives the spectral flux across $k$, $\Pi(k) = -\int_0^k\epsilon_c(k)dk$. While this integral can be carried out analytically, the finite-size contributions will make numerical integration necessary, as discussed below. The energy dissipated at a given wavenumber
is measured by $\epsilon_d(k)$. 

The spectral quantities $\epsilon_{c,d}(k)$ can be measured at a given timestep by explicitly inserting the conservative or dissipative velocities and the vortex positions into 
Eq.~(\ref{eq:fluxes}). One can then average the quantities over different realizations or timesteps.

So far, this calculation does not include the contribution from image vortices in Eq.~(\ref{eq:especsums}). By a time-differentiation of Eq.~(\ref{eq:especsums}), we find that the finite-size correction to the spectral energy flux in Eq.~(\ref{eq:flux}) is 
\begin{widetext}
  \begin{align}
	\dd{E^f(k)}{t} =&\frac\pi k\sum_{l=0}^\infty \epsilon_l J_l(kR) \Bigg\{ \nonumber \\
	  &\sum_i \left[\dot a_{il}\cos(l\theta_i) - la_{il}\dot \theta_i\sin(l\theta_i)\right]\sum_j b_{jl}\cos(l\theta_j) 
	  +\sum_i \left[\dot a_{il}\sin(l\theta_i) + la_{il}\dot \theta_i\cos(l\theta_{ij})\right]\sum_j b_{jl}\sin(l\theta_j) \nonumber\\
	  +&\sum_i a_{il}\cos(l\theta_i)\sum_j \left[\dot b_{jl}\cos(l\theta_j) - lb_{jl}\dot \theta_j\sin(l\theta_j)\right] 
	  +\sum_i a_{il}\sin(l\theta_i)\sum_j \left[\dot b_{jl}\sin(l\theta_j) + lb_{jl}\dot \theta_j\cos(l\theta_j)\right] 
	\Bigg\},\label{eq:dEkdt}
  \end{align}
\end{widetext}
where the remaining derivatives can be found as
\begin{align}
  \dot a_{il} &= 
  la_{il}\frac{\dot r_i}{r_i}, \\
  \dot b_{jl} &= 
  l b_{jl}\frac{\dot r_j}{r_j} + 2kJ_{l+1}(kr_j)\dot r_j, \\
  \dot r_i &= \frac{\vec r_i\cdot \vec v_i}{r_i}, \qquad 
  \dot \theta_i = \frac{\vec r_{i}^\bot\cdot \vec v_{i}}{r_{i}^2}.
\end{align}
Eq.~(\ref{eq:dEkdt}) is still linear in the velocities, so it can be decomposed into a dissipative and a conservative part. However, the products of several Bessel functions makes the integral $\Pi(k) = -\int_0^k \epsilon_c(k)dk$ analytically intractable. One can still measure the time-averaged $\fv{\epsilon_c(k)}$, and then perform the integral numerically by the trapezoidal rule.

The time-averaged spectral energy flux $\fv{\Pi(k)}$ and the spectral energy dissipation rate $\fv{\epsilon_d(k)}$, normalized by the total dissipation rate $\fv{\epsilon_d^{\text{tot}}} = -\fv{\dd{\mathcal{H}}{t}}$ for comparison, are shown for different $\Re$ numbers in Figure~\ref{fig:fluxes}. We see that for sufficiently high Re numbers, the dissipation is mainly localized on the largest scales, while the spectral energy flux is negative in the inertial range corresponding to an inverse energy cascade towards the large scales.

\section{Discussion and conclusions:}
By generalizing Novikov's theory for the energy spectrum of a neutral vortex gas, we find that the Kolmogorov scaling law of the turbulent energy spectrum corresponds to $-4/3$ scaling law of the vorticity correlation function on inertial lengthscales. This is analogous to the Kraichnan-Kolmorogov scaling of the vorticity correlation in classical turbulence.  

In order to explore this statistical correspondence, we propose a novel forced and dissipative point vortex model that is able to produce a statistically steady state turbulent regime, where there is a coexistence between the large-scale Onsager-like vortices of various sizes and the inverse energy cascade. We show that in a system of two-signs vortices, the inverse energy cascade originates from the vortex clustering, that also results in persistent correlations in the vorticity field. Hence, the Kolmogorov $-5/3$ scaling law is directly connected to the scale-free statistics of vorticity, measured by the weighted pair correlation function $g_w(r)\sim r^{-4/3}$. To unify our data obtained for different values of stirring and dissipation parameters, we define a dimensionless $\Re$ number analogously to its classical definition, and find that this $\Re$ for a vortex gas depends solely on the mean vortex number and thermal friction coefficient. 

We study vortex dynamics in a disk similar to the experimental setup of the highly-oblate BEC, so that our results could be compared with future experiments. In particular, with the recent experimental advances on in-situ imaging of vortices in BECs~\cite{wilson2015situ}, it maybe possible to compute the pair correlation function from experimentally obtained vortex configurations and infer the presence of an inverse energy cascade. Because of the boundary effects, we also had to carefully take into account the finite-size corrections. This way, we showed that the vortex-image interactions do not affect the scaling laws, yet they can limit substantially the scaling range, i.e. the spectral gap between the scales where energy is injected and where it is dissipated.

\section*{Acknowledgements:} We are grateful to Nigel Goldenfeld and Yuri Galperin for insightful comments and feedback on the manuscript.

\bibliography{ref}

\begin{thebibliography}{26}%
\makeatletter
\providecommand \@ifxundefined [1]{%
 \@ifx{#1\undefined}
}%
\providecommand \@ifnum [1]{%
 \ifnum #1\expandafter \@firstoftwo
 \else \expandafter \@secondoftwo
 \fi
}%
\providecommand \@ifx [1]{%
 \ifx #1\expandafter \@firstoftwo
 \else \expandafter \@secondoftwo
 \fi
}%
\providecommand \natexlab [1]{#1}%
\providecommand \enquote  [1]{``#1''}%
\providecommand \bibnamefont  [1]{#1}%
\providecommand \bibfnamefont [1]{#1}%
\providecommand \citenamefont [1]{#1}%
\providecommand \href@noop [0]{\@secondoftwo}%
\providecommand \href [0]{\begingroup \@sanitize@url \@href}%
\providecommand \@href[1]{\@@startlink{#1}\@@href}%
\providecommand \@@href[1]{\endgroup#1\@@endlink}%
\providecommand \@sanitize@url [0]{\catcode `\\12\catcode `\$12\catcode
  `\&12\catcode `\#12\catcode `\^12\catcode `\_12\catcode `\%12\relax}%
\providecommand \@@startlink[1]{}%
\providecommand \@@endlink[0]{}%
\providecommand \url  [0]{\begingroup\@sanitize@url \@url }%
\providecommand \@url [1]{\endgroup\@href {#1}{\urlprefix }}%
\providecommand \urlprefix  [0]{URL }%
\providecommand \Eprint [0]{\href }%
\providecommand \doibase [0]{http://dx.doi.org/}%
\providecommand \selectlanguage [0]{\@gobble}%
\providecommand \bibinfo  [0]{\@secondoftwo}%
\providecommand \bibfield  [0]{\@secondoftwo}%
\providecommand \translation [1]{[#1]}%
\providecommand \BibitemOpen [0]{}%
\providecommand \bibitemStop [0]{}%
\providecommand \bibitemNoStop [0]{.\EOS\space}%
\providecommand \EOS [0]{\spacefactor3000\relax}%
\providecommand \BibitemShut  [1]{\csname bibitem#1\endcsname}%
\let\auto@bib@innerbib\@empty
\bibitem [{\citenamefont {Kraichnan}(1967)}]{kraichnan1967inertial}%
  \BibitemOpen
  \bibfield  {author} {\bibinfo {author} {\bibfnamefont {Robert~H}\
  \bibnamefont {Kraichnan}},\ }\bibfield  {title} {\enquote {\bibinfo {title}
  {Inertial ranges in two-dimensional turbulence},}\ }\href@noop {} {\bibfield
  {journal} {\bibinfo  {journal} {Physics of Fluids}\ }\textbf {\bibinfo
  {volume} {10}},\ \bibinfo {pages} {1417--1423} (\bibinfo {year}
  {1967})}\BibitemShut {NoStop}%
\bibitem [{\citenamefont {Neely}\ \emph {et~al.}(2013)\citenamefont {Neely},
  \citenamefont {Bradley}, \citenamefont {Samson}, \citenamefont {Rooney},
  \citenamefont {Wright}, \citenamefont {Law}, \citenamefont
  {Carretero-Gonz{\'a}lez}, \citenamefont {Kevrekidis}, \citenamefont {Davis},\
  and\ \citenamefont {Anderson}}]{neely2013characteristics}%
  \BibitemOpen
  \bibfield  {author} {\bibinfo {author} {\bibfnamefont {T~W}\ \bibnamefont
  {Neely}}, \bibinfo {author} {\bibfnamefont {A~S}\ \bibnamefont {Bradley}},
  \bibinfo {author} {\bibfnamefont {E~C}\ \bibnamefont {Samson}}, \bibinfo
  {author} {\bibfnamefont {S~J}\ \bibnamefont {Rooney}}, \bibinfo {author}
  {\bibfnamefont {E~M}\ \bibnamefont {Wright}}, \bibinfo {author}
  {\bibfnamefont {K~J~H}\ \bibnamefont {Law}}, \bibinfo {author} {\bibfnamefont
  {R}~\bibnamefont {Carretero-Gonz{\'a}lez}}, \bibinfo {author} {\bibfnamefont
  {P~G}\ \bibnamefont {Kevrekidis}}, \bibinfo {author} {\bibfnamefont {M~J}\
  \bibnamefont {Davis}}, \ and\ \bibinfo {author} {\bibfnamefont {B~P}\
  \bibnamefont {Anderson}},\ }\bibfield  {title} {\enquote {\bibinfo {title}
  {Characteristics of two-dimensional quantum turbulence in a compressible
  superfluid},}\ }\href@noop {} {\bibfield  {journal} {\bibinfo  {journal}
  {Physical Review Letters}\ }\textbf {\bibinfo {volume} {111}},\ \bibinfo
  {pages} {235301} (\bibinfo {year} {2013})}\BibitemShut {NoStop}%
\bibitem [{\citenamefont {Reeves}\ \emph {et~al.}(2012)\citenamefont {Reeves},
  \citenamefont {Anderson},\ and\ \citenamefont {Bradley}}]{Reeves_2012}%
  \BibitemOpen
  \bibfield  {author} {\bibinfo {author} {\bibfnamefont {M~T}\ \bibnamefont
  {Reeves}}, \bibinfo {author} {\bibfnamefont {B~P}\ \bibnamefont {Anderson}},
  \ and\ \bibinfo {author} {\bibfnamefont {A~S}\ \bibnamefont {Bradley}},\
  }\bibfield  {title} {\enquote {\bibinfo {title} {Classical and quantum
  regimes of two-dimensional turbulence in trapped bose-einstein
  condensates},}\ }\href@noop {} {\bibfield  {journal} {\bibinfo  {journal}
  {Physical Review A}\ }\textbf {\bibinfo {volume} {86}},\ \bibinfo {pages}
  {053621} (\bibinfo {year} {2012})}\BibitemShut {NoStop}%
\bibitem [{\citenamefont {Bradley}\ and\ \citenamefont
  {Anderson}(2012)}]{Bradley_2012}%
  \BibitemOpen
  \bibfield  {author} {\bibinfo {author} {\bibfnamefont {Ashton~S}\
  \bibnamefont {Bradley}}\ and\ \bibinfo {author} {\bibfnamefont {Brian~P}\
  \bibnamefont {Anderson}},\ }\bibfield  {title} {\enquote {\bibinfo {title}
  {Energy spectra of vortex distributions in two-dimensional quantum
  turbulence},}\ }\href@noop {} {\bibfield  {journal} {\bibinfo  {journal}
  {Physical Review X}\ }\textbf {\bibinfo {volume} {2}},\ \bibinfo {pages}
  {041001} (\bibinfo {year} {2012})}\BibitemShut {NoStop}%
\bibitem [{\citenamefont {Reeves}\ \emph {et~al.}(2013)\citenamefont {Reeves},
  \citenamefont {Billam}, \citenamefont {Anderson},\ and\ \citenamefont
  {Bradley}}]{Reeves_2013}%
  \BibitemOpen
  \bibfield  {author} {\bibinfo {author} {\bibfnamefont {Matthew~T}\
  \bibnamefont {Reeves}}, \bibinfo {author} {\bibfnamefont {Thomas~P}\
  \bibnamefont {Billam}}, \bibinfo {author} {\bibfnamefont {Brian~P}\
  \bibnamefont {Anderson}}, \ and\ \bibinfo {author} {\bibfnamefont {Ashton~S}\
  \bibnamefont {Bradley}},\ }\bibfield  {title} {\enquote {\bibinfo {title}
  {Inverse energy cascade in forced two-dimensional quantum turbulence},}\
  }\href@noop {} {\bibfield  {journal} {\bibinfo  {journal} {Physical review
  letters}\ }\textbf {\bibinfo {volume} {110}},\ \bibinfo {pages} {104501}
  (\bibinfo {year} {2013})}\BibitemShut {NoStop}%
\bibitem [{\citenamefont {Simula}\ \emph {et~al.}(2014)\citenamefont {Simula},
  \citenamefont {Davis},\ and\ \citenamefont {Helmerson}}]{Simula2014}%
  \BibitemOpen
  \bibfield  {author} {\bibinfo {author} {\bibfnamefont {Tapio}\ \bibnamefont
  {Simula}}, \bibinfo {author} {\bibfnamefont {Matthew~J}\ \bibnamefont
  {Davis}}, \ and\ \bibinfo {author} {\bibfnamefont {Kristian}\ \bibnamefont
  {Helmerson}},\ }\bibfield  {title} {\enquote {\bibinfo {title} {Emergence of
  order from turbulence in an isolated planar superfluid},}\ }\href@noop {}
  {\bibfield  {journal} {\bibinfo  {journal} {Physical Review Letters}\
  }\textbf {\bibinfo {volume} {113}},\ \bibinfo {pages} {165302} (\bibinfo
  {year} {2014})}\BibitemShut {NoStop}%
\bibitem [{\citenamefont {Moon}\ \emph {et~al.}(2015)\citenamefont {Moon},
  \citenamefont {Kwon}, \citenamefont {Lee},\ and\ \citenamefont
  {Shin}}]{moon2015thermal}%
  \BibitemOpen
  \bibfield  {author} {\bibinfo {author} {\bibfnamefont {Geol}\ \bibnamefont
  {Moon}}, \bibinfo {author} {\bibfnamefont {Woo~Jin}\ \bibnamefont {Kwon}},
  \bibinfo {author} {\bibfnamefont {Hyunjik}\ \bibnamefont {Lee}}, \ and\
  \bibinfo {author} {\bibfnamefont {Yong-il}\ \bibnamefont {Shin}},\ }\bibfield
   {title} {\enquote {\bibinfo {title} {Thermal friction on quantum vortices in
  a {B}ose-{E}instein condensate},}\ }\href@noop {} {\bibfield  {journal}
  {\bibinfo  {journal} {Physical Review A}\ }\textbf {\bibinfo {volume} {92}},\
  \bibinfo {pages} {051601} (\bibinfo {year} {2015})}\BibitemShut {NoStop}%
\bibitem [{\citenamefont {Billam}\ \emph {et~al.}(2015)\citenamefont {Billam},
  \citenamefont {Reeves},\ and\ \citenamefont {Bradley}}]{billam2015spectral}%
  \BibitemOpen
  \bibfield  {author} {\bibinfo {author} {\bibfnamefont {T~P}\ \bibnamefont
  {Billam}}, \bibinfo {author} {\bibfnamefont {M~T}\ \bibnamefont {Reeves}}, \
  and\ \bibinfo {author} {\bibfnamefont {A~S}\ \bibnamefont {Bradley}},\
  }\bibfield  {title} {\enquote {\bibinfo {title} {Spectral energy transport in
  two-dimensional quantum vortex dynamics},}\ }\href@noop {} {\bibfield
  {journal} {\bibinfo  {journal} {Physical Review A}\ }\textbf {\bibinfo
  {volume} {91}},\ \bibinfo {pages} {023615} (\bibinfo {year}
  {2015})}\BibitemShut {NoStop}%
\bibitem [{\citenamefont {Onsager}(1949)}]{Onsager_1949}%
  \BibitemOpen
  \bibfield  {author} {\bibinfo {author} {\bibfnamefont {Lars}\ \bibnamefont
  {Onsager}},\ }\bibfield  {title} {\enquote {\bibinfo {title} {Statistical
  hydrodynamics},}\ }\href@noop {} {\bibfield  {journal} {\bibinfo  {journal}
  {Il Nuovo Cimento (1943-1954)}\ }\textbf {\bibinfo {volume} {6}},\ \bibinfo
  {pages} {279--287} (\bibinfo {year} {1949})}\BibitemShut {NoStop}%
\bibitem [{\citenamefont {Yu}\ \emph {et~al.}(2016)\citenamefont {Yu},
  \citenamefont {Billam}, \citenamefont {Nian}, \citenamefont {Reeves},\ and\
  \citenamefont {Bradley}}]{yu2016theory}%
  \BibitemOpen
  \bibfield  {author} {\bibinfo {author} {\bibfnamefont {Xiaoquan}\
  \bibnamefont {Yu}}, \bibinfo {author} {\bibfnamefont {Thomas~P}\ \bibnamefont
  {Billam}}, \bibinfo {author} {\bibfnamefont {Jun}\ \bibnamefont {Nian}},
  \bibinfo {author} {\bibfnamefont {Matthew~T}\ \bibnamefont {Reeves}}, \ and\
  \bibinfo {author} {\bibfnamefont {Ashton~S}\ \bibnamefont {Bradley}},\
  }\bibfield  {title} {\enquote {\bibinfo {title} {Theory of the
  vortex-clustering transition in a confined two-dimensional quantum fluid},}\
  }\href@noop {} {\bibfield  {journal} {\bibinfo  {journal} {Physical Review
  A}\ }\textbf {\bibinfo {volume} {94}},\ \bibinfo {pages} {023602} (\bibinfo
  {year} {2016})}\BibitemShut {NoStop}%
\bibitem [{\citenamefont {Skaugen}\ and\ \citenamefont
  {Angheluta}(2016{\natexlab{a}})}]{skaugen2016vortex}%
  \BibitemOpen
  \bibfield  {author} {\bibinfo {author} {\bibfnamefont {Audun}\ \bibnamefont
  {Skaugen}}\ and\ \bibinfo {author} {\bibfnamefont {Luiza}\ \bibnamefont
  {Angheluta}},\ }\bibfield  {title} {\enquote {\bibinfo {title} {Vortex
  clustering and universal scaling laws in two-dimensional quantum
  turbulence},}\ }\href@noop {} {\bibfield  {journal} {\bibinfo  {journal}
  {Physical Review E}\ }\textbf {\bibinfo {volume} {93}},\ \bibinfo {pages}
  {032106} (\bibinfo {year} {2016}{\natexlab{a}})}\BibitemShut {NoStop}%
\bibitem [{\citenamefont {Novikov}(1975)}]{Novikov_1975}%
  \BibitemOpen
  \bibfield  {author} {\bibinfo {author} {\bibfnamefont {EA}~\bibnamefont
  {Novikov}},\ }\bibfield  {title} {\enquote {\bibinfo {title} {Dynamics and
  statistics of a system of vortices},}\ }\href@noop {} {\bibfield  {journal}
  {\bibinfo  {journal} {Zh. Eksp. Teor. Fiz}\ }\textbf {\bibinfo {volume}
  {68}},\ \bibinfo {pages} {1868--1882} (\bibinfo {year} {1975})}\BibitemShut
  {NoStop}%
\bibitem [{\citenamefont {Skaugen}\ and\ \citenamefont
  {Angheluta}(2016{\natexlab{b}})}]{skaugen2016velocity}%
  \BibitemOpen
  \bibfield  {author} {\bibinfo {author} {\bibfnamefont {Audun}\ \bibnamefont
  {Skaugen}}\ and\ \bibinfo {author} {\bibfnamefont {Luiza}\ \bibnamefont
  {Angheluta}},\ }\bibfield  {title} {\enquote {\bibinfo {title} {Velocity
  statistics for nonuniform configurations of point vortices},}\ }\href@noop {}
  {\bibfield  {journal} {\bibinfo  {journal} {Physical Review E}\ }\textbf
  {\bibinfo {volume} {93}},\ \bibinfo {pages} {042137} (\bibinfo {year}
  {2016}{\natexlab{b}})}\BibitemShut {NoStop}%
\bibitem [{\citenamefont {Pismen}(1999)}]{pismen1999}%
  \BibitemOpen
  \bibfield  {author} {\bibinfo {author} {\bibfnamefont {Len~M}\ \bibnamefont
  {Pismen}},\ }\href@noop {} {\emph {\bibinfo {title} {Vortices in nonlinear
  fields: From liquid crystals to superfluids, from non-equilibrium patterns to
  cosmic strings}}},\ Vol.\ \bibinfo {volume} {100}\ (\bibinfo  {publisher}
  {Oxford University Press},\ \bibinfo {year} {1999})\BibitemShut {NoStop}%
\bibitem [{\citenamefont {Newton}(2013)}]{newton2013n}%
  \BibitemOpen
  \bibfield  {author} {\bibinfo {author} {\bibfnamefont {Paul~K}\ \bibnamefont
  {Newton}},\ }\href@noop {} {\emph {\bibinfo {title} {The {N}-vortex problem:
  analytical techniques}}},\ Vol.\ \bibinfo {volume} {145}\ (\bibinfo
  {publisher} {Springer Science \& Business Media},\ \bibinfo {year}
  {2013})\BibitemShut {NoStop}%
\bibitem [{\citenamefont {Benzi}\ \emph {et~al.}(1992)\citenamefont {Benzi},
  \citenamefont {Colella}, \citenamefont {Briscolini},\ and\ \citenamefont
  {Santangelo}}]{benzi1992simple}%
  \BibitemOpen
  \bibfield  {author} {\bibinfo {author} {\bibfnamefont {R}~\bibnamefont
  {Benzi}}, \bibinfo {author} {\bibfnamefont {M}~\bibnamefont {Colella}},
  \bibinfo {author} {\bibfnamefont {M}~\bibnamefont {Briscolini}}, \ and\
  \bibinfo {author} {\bibfnamefont {P}~\bibnamefont {Santangelo}},\ }\bibfield
  {title} {\enquote {\bibinfo {title} {A simple point vortex model for
  two-dimensional decaying turbulence},}\ }\href@noop {} {\bibfield  {journal}
  {\bibinfo  {journal} {Physics of Fluids A: Fluid Dynamics (1989-1993)}\
  }\textbf {\bibinfo {volume} {4}},\ \bibinfo {pages} {1036--1039} (\bibinfo
  {year} {1992})}\BibitemShut {NoStop}%
\bibitem [{\citenamefont {Pointin}\ and\ \citenamefont
  {Lundgren}(1976)}]{pointin1976statistical}%
  \BibitemOpen
  \bibfield  {author} {\bibinfo {author} {\bibfnamefont {Yves~Bernard}\
  \bibnamefont {Pointin}}\ and\ \bibinfo {author} {\bibfnamefont
  {TS}~\bibnamefont {Lundgren}},\ }\bibfield  {title} {\enquote {\bibinfo
  {title} {Statistical mechanics of two-dimensional vortices in a bounded
  container},}\ }\href@noop {} {\bibfield  {journal} {\bibinfo  {journal}
  {Physics of Fluids (1958-1988)}\ }\textbf {\bibinfo {volume} {19}},\ \bibinfo
  {pages} {1459--1470} (\bibinfo {year} {1976})}\BibitemShut {NoStop}%
\bibitem [{\citenamefont {Valani}\ \emph {et~al.}(2016)\citenamefont {Valani},
  \citenamefont {Groszek},\ and\ \citenamefont {Simula}}]{valani2016einstein}%
  \BibitemOpen
  \bibfield  {author} {\bibinfo {author} {\bibfnamefont {Rahil~N}\ \bibnamefont
  {Valani}}, \bibinfo {author} {\bibfnamefont {Andrew~J}\ \bibnamefont
  {Groszek}}, \ and\ \bibinfo {author} {\bibfnamefont {Tapio~P}\ \bibnamefont
  {Simula}},\ }\bibfield  {title} {\enquote {\bibinfo {title} {Einstein-{B}ose
  condensation of {O}nsager {V}ortices},}\ }\href@noop {} {\bibfield  {journal}
  {\bibinfo  {journal} {arXiv preprint arXiv:1612.02930}\ } (\bibinfo {year}
  {2016})}\BibitemShut {NoStop}%
\bibitem [{\citenamefont {Kim}\ \emph {et~al.}(2016)\citenamefont {Kim},
  \citenamefont {Kwon},\ and\ \citenamefont {Shin}}]{kim2016role}%
  \BibitemOpen
  \bibfield  {author} {\bibinfo {author} {\bibfnamefont {Joon~Hyun}\
  \bibnamefont {Kim}}, \bibinfo {author} {\bibfnamefont {Woo~Jin}\ \bibnamefont
  {Kwon}}, \ and\ \bibinfo {author} {\bibfnamefont {Y}~\bibnamefont {Shin}},\
  }\bibfield  {title} {\enquote {\bibinfo {title} {Role of thermal friction in
  relaxation of turbulent {B}ose-{E}instein condensates},}\ }\href@noop {}
  {\bibfield  {journal} {\bibinfo  {journal} {Physical Review A}\ }\textbf
  {\bibinfo {volume} {94}},\ \bibinfo {pages} {033612} (\bibinfo {year}
  {2016})}\BibitemShut {NoStop}%
\bibitem [{\citenamefont {Campbell}\ and\ \citenamefont
  {O'Neil}(1991)}]{campbell1991statistics}%
  \BibitemOpen
  \bibfield  {author} {\bibinfo {author} {\bibfnamefont {LJ}~\bibnamefont
  {Campbell}}\ and\ \bibinfo {author} {\bibfnamefont {Kevin}\ \bibnamefont
  {O'Neil}},\ }\bibfield  {title} {\enquote {\bibinfo {title} {Statistics of
  two-dimensional point vortices and high-energy vortex states},}\ }\href@noop
  {} {\bibfield  {journal} {\bibinfo  {journal} {Journal of Statistical
  Physics}\ }\textbf {\bibinfo {volume} {65}},\ \bibinfo {pages} {495--529}
  (\bibinfo {year} {1991})}\BibitemShut {NoStop}%
\bibitem [{\citenamefont {Siggia}\ and\ \citenamefont {Aref}(1981)}]{Siggia}%
  \BibitemOpen
  \bibfield  {author} {\bibinfo {author} {\bibfnamefont {Eric~D}\ \bibnamefont
  {Siggia}}\ and\ \bibinfo {author} {\bibfnamefont {Hassan}\ \bibnamefont
  {Aref}},\ }\bibfield  {title} {\enquote {\bibinfo {title} {Point-vortex
  simulation of the inverse energy cascade in two-dimensional turbulence},}\
  }\href@noop {} {\bibfield  {journal} {\bibinfo  {journal} {Physics of Fluids
  (1958-1988)}\ }\textbf {\bibinfo {volume} {24}},\ \bibinfo {pages} {171--173}
  (\bibinfo {year} {1981})}\BibitemShut {NoStop}%
\bibitem [{sup()}]{supplemental}%
  \BibitemOpen
  \href@noop {} {}\bibinfo {note} {See the supplementary material for two
  movies illustrating the vortex dynamics during the buildup towards
  turbulence, and the statistically stationary turbulent regime,
  respectively.}\BibitemShut {Stop}%
\bibitem [{\citenamefont {McLachlan}\ and\ \citenamefont
  {Atela}(1992)}]{mclachlan1992accuracy}%
  \BibitemOpen
  \bibfield  {author} {\bibinfo {author} {\bibfnamefont {Robert~I}\
  \bibnamefont {McLachlan}}\ and\ \bibinfo {author} {\bibfnamefont {Pau}\
  \bibnamefont {Atela}},\ }\bibfield  {title} {\enquote {\bibinfo {title} {The
  accuracy of symplectic integrators},}\ }\href@noop {} {\bibfield  {journal}
  {\bibinfo  {journal} {Nonlinearity}\ }\textbf {\bibinfo {volume} {5}},\
  \bibinfo {pages} {541} (\bibinfo {year} {1992})}\BibitemShut {NoStop}%
\bibitem [{\citenamefont {Groszek}\ \emph {et~al.}(2016)\citenamefont
  {Groszek}, \citenamefont {Simula}, \citenamefont {Paganin},\ and\
  \citenamefont {Helmerson}}]{groszek2016onsager}%
  \BibitemOpen
  \bibfield  {author} {\bibinfo {author} {\bibfnamefont {Andrew~J}\
  \bibnamefont {Groszek}}, \bibinfo {author} {\bibfnamefont {Tapio~P}\
  \bibnamefont {Simula}}, \bibinfo {author} {\bibfnamefont {David~M}\
  \bibnamefont {Paganin}}, \ and\ \bibinfo {author} {\bibfnamefont {Kristian}\
  \bibnamefont {Helmerson}},\ }\bibfield  {title} {\enquote {\bibinfo {title}
  {Onsager vortex formation in {B}ose-{E}instein condensates in two-dimensional
  power-law traps},}\ }\href@noop {} {\bibfield  {journal} {\bibinfo  {journal}
  {Physical Review A}\ }\textbf {\bibinfo {volume} {93}},\ \bibinfo {pages}
  {043614} (\bibinfo {year} {2016})}\BibitemShut {NoStop}%
\bibitem [{\citenamefont {Yoshida}\ and\ \citenamefont
  {M.~Sano}(2005)}]{yoshida2005numerical}%
  \BibitemOpen
  \bibfield  {author} {\bibinfo {author} {\bibfnamefont {Takeshi}\ \bibnamefont
  {Yoshida}}\ and\ \bibinfo {author} {\bibfnamefont {Mitsusada}\ \bibnamefont
  {M.~Sano}},\ }\bibfield  {title} {\enquote {\bibinfo {title} {Numerical
  simulation of vortex crystals and merging in {N}-point vortex systems with
  circular boundary},}\ }\href@noop {} {\bibfield  {journal} {\bibinfo
  {journal} {Journal of the Physical Society of Japan}\ }\textbf {\bibinfo
  {volume} {74}},\ \bibinfo {pages} {587--598} (\bibinfo {year}
  {2005})}\BibitemShut {NoStop}%
\bibitem [{\citenamefont {Wilson}\ \emph {et~al.}(2015)\citenamefont {Wilson},
  \citenamefont {Newman}, \citenamefont {Lowney},\ and\ \citenamefont
  {Anderson}}]{wilson2015situ}%
  \BibitemOpen
  \bibfield  {author} {\bibinfo {author} {\bibfnamefont {Kali~E}\ \bibnamefont
  {Wilson}}, \bibinfo {author} {\bibfnamefont {Zachary~L}\ \bibnamefont
  {Newman}}, \bibinfo {author} {\bibfnamefont {Joseph~D}\ \bibnamefont
  {Lowney}}, \ and\ \bibinfo {author} {\bibfnamefont {Brian~P}\ \bibnamefont
  {Anderson}},\ }\bibfield  {title} {\enquote {\bibinfo {title} {In situ
  imaging of vortices in {B}ose-{E}instein condensates},}\ }\href@noop {}
  {\bibfield  {journal} {\bibinfo  {journal} {Physical Review A}\ }\textbf
  {\bibinfo {volume} {91}},\ \bibinfo {pages} {023621} (\bibinfo {year}
  {2015})}\BibitemShut {NoStop}%
\end{thebibliography}%

\end{document}